\documentclass[
floatfix,
aip,
apl,
twocolumn,
superscriptaddress,
amssymb,
 groupaddress,
]{revtex4}

\usepackage{float}
\usepackage{amssymb}
\usepackage{gensymb}
\usepackage{latexsym}
\usepackage[pdftex]{graphicx}
\usepackage{amsmath}
\usepackage{graphicx}
\usepackage{dcolumn}
\usepackage{amsfonts}
\usepackage{bm}
\usepackage{epsfig}
\usepackage{siunitx}

\begin{document}
\title{Toward durable Al-InSb hybrid heterostructures via epitaxy of 2ML interfacial InAs screening layers}

\author{C. Thomas}
\affiliation{Department of Physics and Astronomy, Purdue University, West Lafayette, IN 47907, USA}
\affiliation{Birck Nanotechnology Center, Purdue University, West Lafayette, IN 47907 USA}

\author{R. E. Diaz}
\affiliation{Birck Nanotechnology Center, Purdue University, West Lafayette, IN 47907 USA}

\author{J. H. Dycus}
\affiliation{Eurofins EAG Materials Science, Raleigh, NC 27606 USA}

\author{M. E. Salmon}
\affiliation{Eurofins EAG Materials Science, Raleigh, NC 27606 USA}

\author{R. E. Daniel}
\affiliation{Eurofins EAG Materials Science, Raleigh, NC 27606 USA}

\author{T. Wang}
\affiliation{Department of Physics and Astronomy, Purdue University, West Lafayette, IN 47907, USA}
\affiliation{Birck Nanotechnology Center, Purdue University, West Lafayette, IN 47907 USA}

\author{G. C. Gardner}
\affiliation{Birck Nanotechnology Center, Purdue University, West Lafayette, IN 47907 USA}
\affiliation{Microsoft Quantum Purdue, Purdue University, West Lafayette, IN 47907, USA}

\author{M. J. Manfra}
\affiliation{Department of Physics and Astronomy, Purdue University, West Lafayette, IN 47907, USA}
\affiliation{Birck Nanotechnology Center, Purdue University, West Lafayette, IN 47907 USA}
\affiliation{Microsoft Quantum Purdue, Purdue University, West Lafayette, IN 47907, USA}
\affiliation{School of Electrical and Computer Engineering, Purdue University, West Lafayette, IN 47907, USA}
\affiliation{School of Materials Engineering, Purdue University, West Lafayette, IN 47907 USA}

\received{\today}

\begin{abstract}
The large Land\'{e} g-factor, high spin-orbit coupling, and low effective mass of the two-dimensional electron gas in InSb quantum wells combined with proximal superconductivity may realize a scalable platform for topological quantum computation. Aluminum thin films directly deposited on top of InSb planar structures result in the formation of a reactive AlInSb layer at the interface. This interlayer progressively consumes the whole Al film, resulting in a disordered AlInSb layer after few months at room temperature.  
We report on a heterostructure design that results in a significant increase of the durability of these hybrid Al-InSb heterostructures with the preservation of a pure Al film and sharp superconductor-semiconductor interface for more than one year. Two monolayers of epitaxial InAs at the superconductor-semiconductor interface prevent interfacial reactivity as evidenced by X-ray reflectivity and energy dispersive spectroscopy measurements. Structural characterizations of the Al films by transmission electron microscopy reveal the presence of tens of nanometers wide grains predominantly oriented with Al(110) parallel to InSb(001).

\end{abstract} 

\maketitle

\section{Introduction}
Hybrid heterostructures of semiconductors with high spin-orbit coupling and s-wave superconductors are expected to be topological superconductors hosting Majorana zero modes (MZMs) upon application of an in-plane magnetic field \citep{Lutchyn2010} \cite{Sau2010} \citep{Oreg2010} \citep{Alicea2010}. Non-abelian MZMs allow encoding information non-locally, thus forming the building blocks of topological quantum computation. With large spin-orbit coupling, high Land\'{e} g factor and low effective mass, InSb is an attractive semiconductor platform for generation of MZMs \citep{Ke2019}. \textit{In-situ} epitaxial deposition of superconducting materials such as Al has been studied and developed on various semiconductor materials such as Si \citep{Strongin1973}, GaAs \citep{Ludeke1973} \citep{Tournet2016} and more recently on high spin-orbit coupled nanowires of InAs \citep{Krogstrup2015} and InSb \citep{Gazibegovic2017}, demonstrating high quality and low disordered interfaces \citep{Chang2015}.

Deposition of Al films on InSb has been studied previously, for example, with the investigation of room temperature evaporation of Al on InSb (110) films by \textit{in-situ} photoemission analysis \citep{Boscherini1987} \citep{Pei1988}. The formation of AlInSb or AlSb at the interface, accompanied by In clusters on the surface,  has been reported after the deposition of only few Angstroms of Al. 
These photoemission results suggest that the reaction is dominated by Al-In exchange at the superconductor-semiconductor interface \citep{Pei1988}.

Recently, the possibility to generate MZMs with Al-InSb heterostructures has motivated new growth investigations. The growth of Al on InSb nanowires has been reported with Al epitaxially deposited at a low temperature of 120 K on oxide-free InSb nanowires after an atomic hydrogen surface cleaning \citep{Gazibegovic2017} \citep{Zhang2018}. The ability to induce superconductivity between the Al film and the InSb nanowire has been evidenced through the generation of a hard gap of 0.24 meV \citep{Gazibegovic2017}. However, the durability of these hybrid nanowires was not discussed. 

Here we present data on the longevity of \textit{in-situ} epitaxially deposited Al layers on planar InSb (001) structures. The direct epitaxy of Al on InSb  leads to the formation of a reactive AlInSb layer that quickly begins to degrade the interface. Measurable partial consumption of the Al layer is observed within 2 months stored under a nitrogen atmosphere at room temperature. Within 210 days, the entire Al film has been consumed and replaced by a disordered Al$_{0.8}$In$_{0.2}$Sb layer. 
We show that the incorporation of a 2 monolayers (ML) thick InAs screening layer significantly mitigates this effect, allowing preservation of a pure Al layer for more than 390 days at room temperature.
The chemical and structural properties of the Al layers deposited on top of 2ML InAs/InSb heterostructures are presented. 20-30 nm wide Al grains are observed and are predominantly oriented with Al (110) parallel to InSb (001).

\section{Molecular beam epitaxy growth}
\begin{figure}[h]
\includegraphics[scale=0.55]{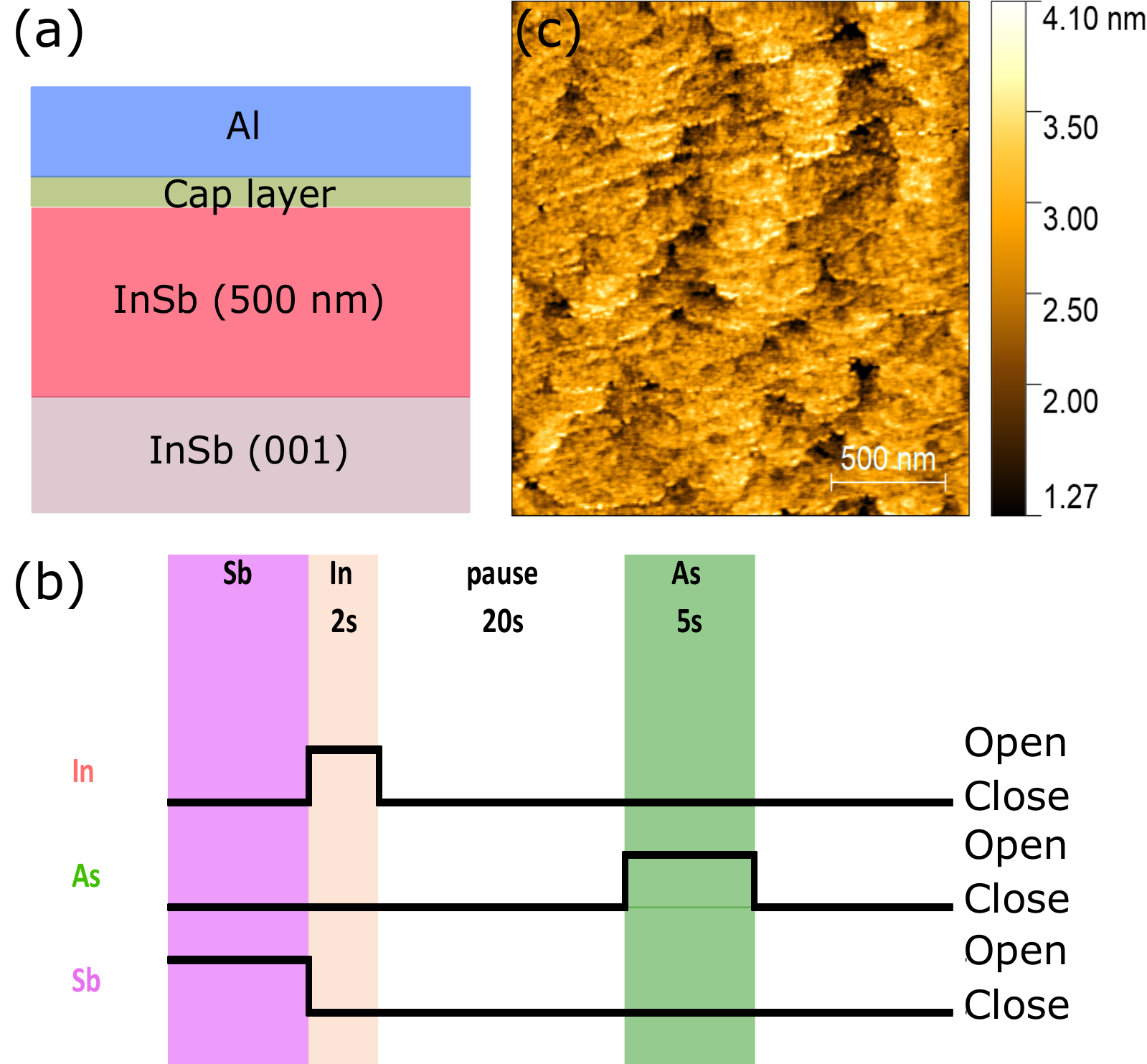} 
\caption{(a) Typical layer stack of the three samples considered in this study. The cap layer is 1ML In for sample A, 2ML InAs for sample B and 40 nm In$_{0.9}$Al$_{0.1}$Sb followed by 2ML InAs for sample C. 
(b) Shutter sequence used for the transition and growth of the first ML of InAs screening layer by migration enhanced epitaxy for samples B and C. (c) $2 \times 2 \mu m^2$ atomic force micrograph of sample B. Root mean square roughness is 0.4 nm. }
\label{Figure1}
\end{figure}

InSb-based heterostructures were grown on InSb (001) substrates by molecular beam epitaxy in a Veeco 930 using ultra-high purity techniques and methods as described in \citep{Gardner2016}. Substrate temperature was measured by blackbody radiation emission. The native oxide of the substrate was removed using atomic hydrogen at 250 \degree C  with a hydrogen pressure of about $3\times 10^{-7}$ T for 30 minutes. This process was done without Sb flux on the surface and resulted in a (4x2) surface reconstruction, indicating an In stabilized surface \citep{Weiss2007}.  
The substrate temperature was then raised to 440 \degree C under Sb flux to grow a 500 nm-thick InSb buffer layer.   Temperature calibration was accomplished by referencing the c(4x4)/p(1x3) reconstruction transition occurring at 400 \degree C for Sb/In ratio $\sim 5$ \citep{Oe1980}, which is used here for the growth of InSb.  
To understand the interaction between the Al films and the Sb-based layers, InSb structures were covered with different cap layers. Three different structures were grown with the layer stacks presented in Figure \ref{Figure1}(a). Sample A ends with 1ML of In to obtain a similar surface reconstruction as the hydrogen cleaned nanowires studied in \citep{Gazibegovic2017} and \citep{Zhang2018}, sample B is terminated with 2ML of InAs  and sample C is completed with a 40 nm layer of In$_{0.9}$Al$_{0.1}$Sb followed by 2ML of InAs. 
With 40 nm of In$_{0.9}$Al$_{0.1}$Sb on top of InSb buffer, sample C enables to study Al epitaxy on a surface that mimics the InSb/In$_{0.9}$Al$_{0.1}$Sb quantum well heterostructures widely reported in literature \citep{Lehner2018} \citep{Yi2015} \citep{Goldammer1998}. 

 The choice of InAs as an Sb-free screening layer is justified by the numerous investigations of Al epitaxy on InAs and the stability of the resulting interface \citep{Krogstrup2015} \citep{Shabani2016}. Moreover, the low bandgap of InAs and the resulting accumulation layer  at Al-InAs interface are known to favor induced superconductivity \citep{Mikkelsen2018} \citep{Antipov2018}. 
The transition to InAs was accomplished using a shutter sequence described in \citep{Tuttle1990}. Because of its large lattice mismatch with InSb ($- 6.4\%$) and to reduce the formation of defects, the InAs screening layer was grown by migration enhanced epitaxy, with the shutter sequence presented in Figure \ref{Figure1}(b), at a low temperature of $350 \degree$C and its thickness was limited to 2ML.

The superconducting Al layer was evaporated from an effusion cell in a Veeco 620 chamber connected under ultra high vacuum to the main III-V growth chamber where semiconductor epitaxy took place.  In addition to the traditional cryopanels mounted to the chamber walls, a smaller moveable cryocooler is used to contact and cool the wafer.  This container, which maintains continuous liquid nitrogen (LN$_2$) flow, has a surface designed to contact the substrate carrier around the perimeter of the wafer without damaging the pristine semiconductor surface.  It is necessary to directly touch the substrate carrier due to extremely low radiative coupling at low temperatures in a vacuum environment. The face of the cooler, which is in contact with the substrate carrier, is made of silver coated copper to aid thermal conduction.  One type K thermocouple is mounted to the cooler and another spring loaded thermocouple touches the substrate.  This system allows a substrate temperature approaching that of LN$_2$ to be reached within few hours.

Al was deposited on samples A, B and C with the same conditions of temperature and growth rate (2 nm/min, calibrated by an \textit{in-situ} quartz crystal microbalance). Samples A and B were covered with 20 nm of Al, while only 7 nm was deposited on sample C.
Immediately after the Al deposition, the samples were moved to a different chamber and oxidized in a controlled manner for 15 min under a O$_2$ pressure of $5\times 10^{-5}$T to stabilize the Al films  \citep{Gazibegovic2017}. 
The samples discussed in this letter were kept at room temperature in a dry environment under nitrogen flow between experiments. 
 
Surface morphology after Al deposition was characterized by atomic force microscopy in tapping mode. Figure \ref{Figure1}(c) displays a 2x2 $\mu$m$^2$ micrograph of sample B. Similar morphology was observed for the other samples.  Surface morphology is identical to the bare semiconductor, indicating a high quality, uniform and conformal Al film. The surface is characterized by clear atomic steps and a low roughness of 0.4 nm.

\section{Results}
\subsection{Aging of Al-InSb structures}
Preparation of InSb-based samples for characterization techniques such as transmission electron microscopy (TEM) or into devices for low temperature electrical measurements is made difficult by the rapid degradation of this material system with application of heat. The addition of Al makes these processes even more complex with potential reaction between Al and InSb-based layers even at room temperature  \citep{Boscherini1987}  \citep{Pei1988}. 
To isolate intrinsic sample properties from those generated by processing degradation, we first study our structures with a non-destructive characterization method, X-ray reflectivity (XRR), requiring no sample preparation. XRR provides useful information on the thickness of the structure top layers with the oscillation periodicity. The slope of the spectrum reflects the top surface roughness while the damping of the oscillations informs on the superconductor-semiconductor interface roughness.  More importantly for our study, the critical angle of reflection gives information on the top layer density and the oscillation amplitude indicates the density difference between the top layers, here nominally Al and InSb. We use the density values extracted from XRR spectra to assess the purity of the Al layer.

\begin{figure*}[h]
\includegraphics[scale=0.5]{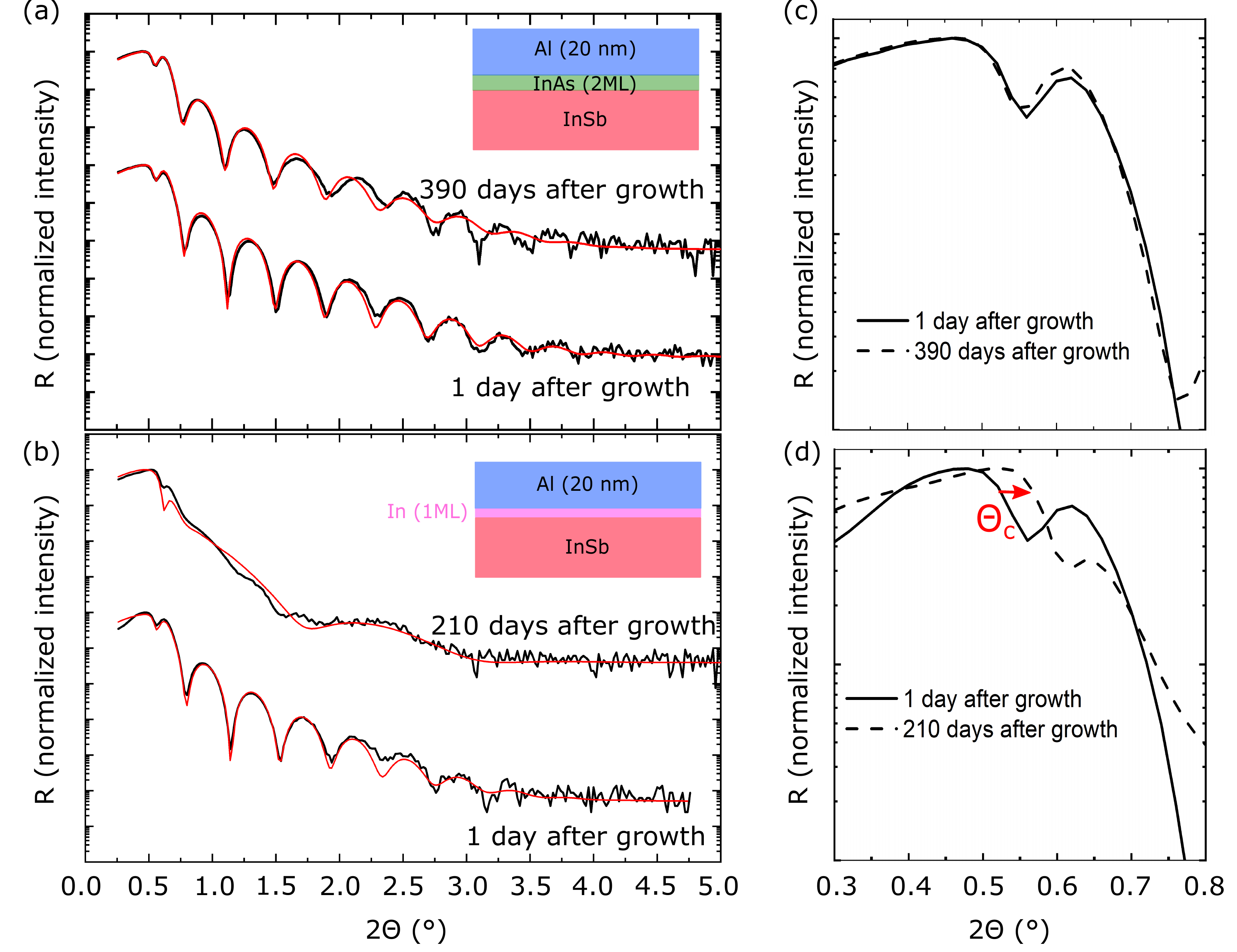} 
\caption{ Xray Reflectivity (XRR) spectra with normalized reflectivity R as a function of $2\theta$  for (a) sample B, one day and 390 days  (shifted for clarity) after growth, and (b) sample A, one day and 210 days (shifted for clarity) after growth. Insets of (a) and (b) represent the nominal layer stack of samples B and A, respectively. 
The red curves correspond to the fits using X'pert Reflectivity \citep{Reflectivity}. A zoom-in at low angle is provided in (c) and (d) for XRR spectra of sample B and sample A, respectively, to see the evolution of critical angle $\theta_c$ and thus density of the top layer with time.   }
\label{Figure2}
\end{figure*}

XRR measurements were performed using a X'pert PANalyical diffractometer with a copper X-ray tube operating at a wavelength $\lambda = \SI{1.5406}{\angstrom}$. Figures \ref{Figure2}(a) and (b) report reflectivity (R) spectra acquired one day and more than 200 days after the growth for samples B and A, respectively.  Note that this study has also been performed on sample C, giving similar results as sample B. The XRR spectra presented in this study have been fitted using PANalytical X'pert Reflectivity software \citep{Reflectivity}.

Immediately after growth, the XRR spectra of samples A and B are very similar  and can be fitted (see red curves) with $19.8 \pm 0.2$ nm of Al (density 2.70 g/cm$^3$) and $1.7 \pm 0.3$ nm of Al$_2$O$_3$ (density 3.95 g/cm$^3$)  on top of the InSb-based semiconductor structure (density of InSb 5.78 g/cm$^3$).
Identical measurements were performed at different time intervals after the growth. The oscillation amplitude of XRR spectra for sample A started decaying progressively within two months after growth  until being barely resolved after 210 days as can be seen in Figure \ref{Figure2} (b).  This large decay of the oscillation amplitude indicates a significant reduction of the density difference between the two top layers and is consistent with the complete transformation of the Al layer into an inhomogeneous AlInSb compound. This progressive transformation of the sample was also visible to the naked eye by a change of color of the sample surface. 
The XRR spectrum acquired 210 days after growth for sample A can be fitted (see red curve) with a
$\sim 5$ nm-thick high density In-rich layer at the interface directly followed by 20 nm of Al$_{0.8}$In$_{0.2}$Sb and 2 nm of Al$_2$O$_3$. 
This is corroborated by a drift of the critical angle $\theta_c$ toward larger angles as emphasized by the red arrow in Figure \ref{Figure2}(d). 
Indeed, $\theta_c$, which corresponds to the maximum angle value that leads to total reflection (R=1), is directly related to the top surface material density $\rho$ by $\theta_c= \sqrt{2 \delta}$ where $\delta = \lambda^2 \frac{N_a r_0}{2\pi} \sum_j \frac{\rho_j}{M_j} (Z_j - f_j^{'})$ \citep{Parratt1954} \citep{Tanner2018} is the dispersion term of the refractive index $n= 1- \delta +i \beta$, $r_0$ is the classical electron radius \citep{note_electron_radius} and $N_a$ is Avogadro's number. $\rho_j$, $M_j$, $Z_j$ and $f_j^{'}$ are the density, mass number, atomic number and real part of the dispersion correction of element $j$ in the considered material, respectively. 
 Just after growth, $\theta_c$ value is in agreement with the density of pure Al, while after 210 days, it indicates a top surface density of 4.55 g/cm$^3$ corresponding to Al$_{0.8}$In$_{0.2}$Sb. 
 
The same experiment was performed on sample B as shown in Figure \ref{Figure2}(a). The XRR spectrum acquired 390 days after the growth is similar to the one obtained just after the growth and can be fitted with 19.7 nm of pure Al covered with 1.9 nm of Al$_2$O$_3$. Consistently, the critical angle value does not change as evidenced in Figure \ref{Figure2}(c). These data validate the use of 2ML InAs screening layer to mitigate the intermixing between the Al films and the InSb-based layers underneath. A significant preservation of the hybrid heterostructure is demonstrated. It is worth noting that the monitoring by XRR of Al quality of sample B is still ongoing (see Section A of Supplementary Information).

\subsection{Chemical and structural characterizations by scanning transmission electron microscopy}
Complementary to the XRR analysis, we performed scanning transmission electron microscopy (STEM) with energy dispersive X-ray spectroscopy (EDX) to characterize the composition of the Al layer and the chemical sharpness of the superconductor-semiconductor interface for sample C.  

TEM sample preparation was performed in a FEI Helios 660 Dual Beam Focused Ion Beam (FIB) system. Sample C was covered with a protective carbon coating prior FIB lift-out and then thinned down to electron transparency at room temperature under a low beam voltage of 2kV. 
Such low voltage  mitigates In droplet formation, reduces the number of defects generated by the FIB preparation close to the superconductor-semiconductor interface, and prevents the diffusion of In and Sb in the Al layer, which was observed for higher beam voltages. Despite all these improvements, the sample preparation technique may still be optimized further. It can not be excluded that some of the defects observed on the presented micrographs are due to the preparation itself. 
STEM imaging was performed using a probe aberration corrected Hitachi 2700 STEM system with an accelerating voltage of 200 kV and a nominal probe size of around 0.1 nm. Observation was made along the $\left<110 \right>$ zone axis in the semiconductor. 

\begin{figure}[h]
\includegraphics[scale=1]{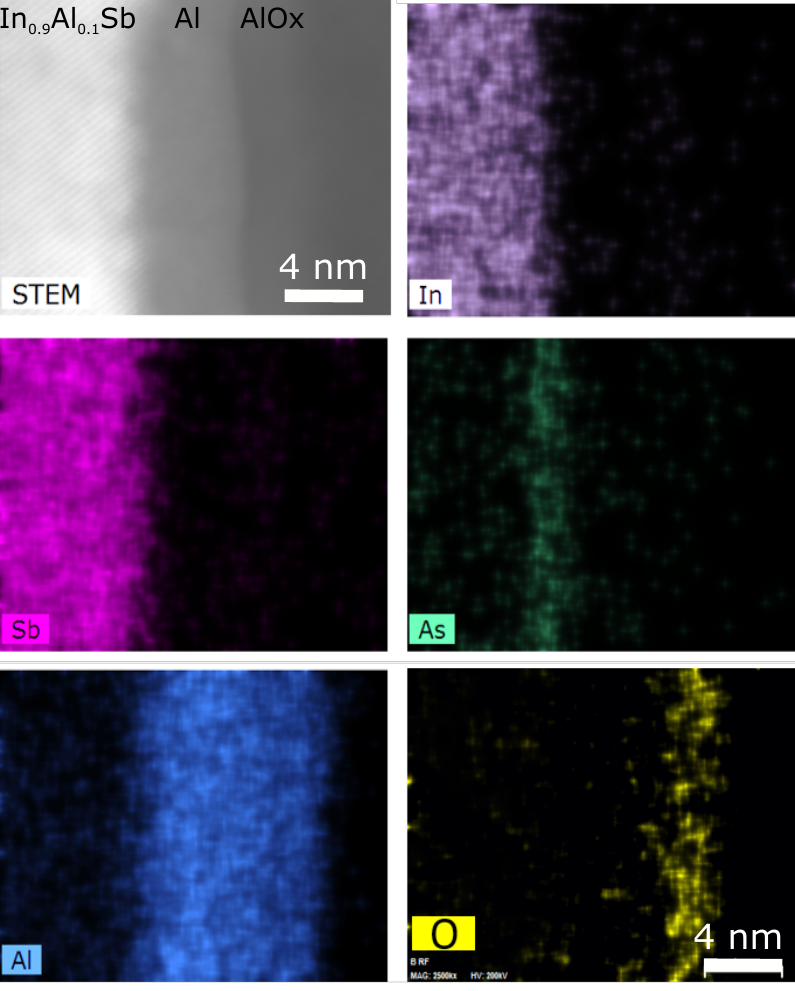} 
\caption{Scanning Transmission Electron Microscopy (STEM) with Energy Dispersive X-ray Spectroscopy (EDX) of the Al-In$_{0.9}$Al$_{0.1}$Sb interface of sample C at room temperature. Left top corner image is the Z-contrast STEM image of the probed region. The other quadrants corresponds to the chemical maps of In (in purple), Sb (in magenta), As (in green), Al (in blue) and oxygen (in yellow), acquired using In-L, Sb-L, As-L, Al-K, and O-K lines, respectively.   }
\label{Figure3}
\end{figure}

Figure \ref{Figure3} displays the EDX analysis performed on sample C with the chemical maps of Sb, In, As, Al and oxygen. Sharp interfaces are observable between the different layers of In$_{0.9}$Al$_{0.1}$Sb, InAs  and Al. A thin layer of aluminum oxide covers the structure. The As chemical signal of the screening layer clearly marks the interface between the Al layer and Sb-based semiconductor underneath. Sb and In signals stop at this interface with no evidence of interdiffusion, confirming the role of the 2ML InAs screening layer and the quality of sample preparation.  \\

The large lattice mismatch between Al (\SI{4.05}{\angstrom}) and InSb (\SI{6.479}{\angstrom}) results in the formation of interfacial domains to reduce the mismatch and the strain at the superconductor-semiconductor interface \citep{Zheleva1994}, similarly to what has been observed for Al-InAs hybrid structures \citep{Krogstrup2015}. These domains align a number $n_f$ of lattice parameters $a_f$ of the Al film with a number $n_s$ of lattice constants $a_s$ of the semiconductor underneath. The associated mismatch is calculated by $\frac{n_f a_f - n_s a_s}{n_s a_s}$ and is estimated to be of few $\%$ between Al and InSb (\textit{e.g.} $4 \%$ for Al (001) growth on InSb (001) with $n_f=5$ and $n_s=3$ \citep{Krogstrup2015}). This large domain mismatch and the important lattice mismatch between InAs screening layers and InSb ($-6.4 \%$) motivate the structural characterizations of the epitaxially deposited Al films and of their interface with the semiconductors.

\begin{figure*}[h]
\includegraphics[scale=0.9]{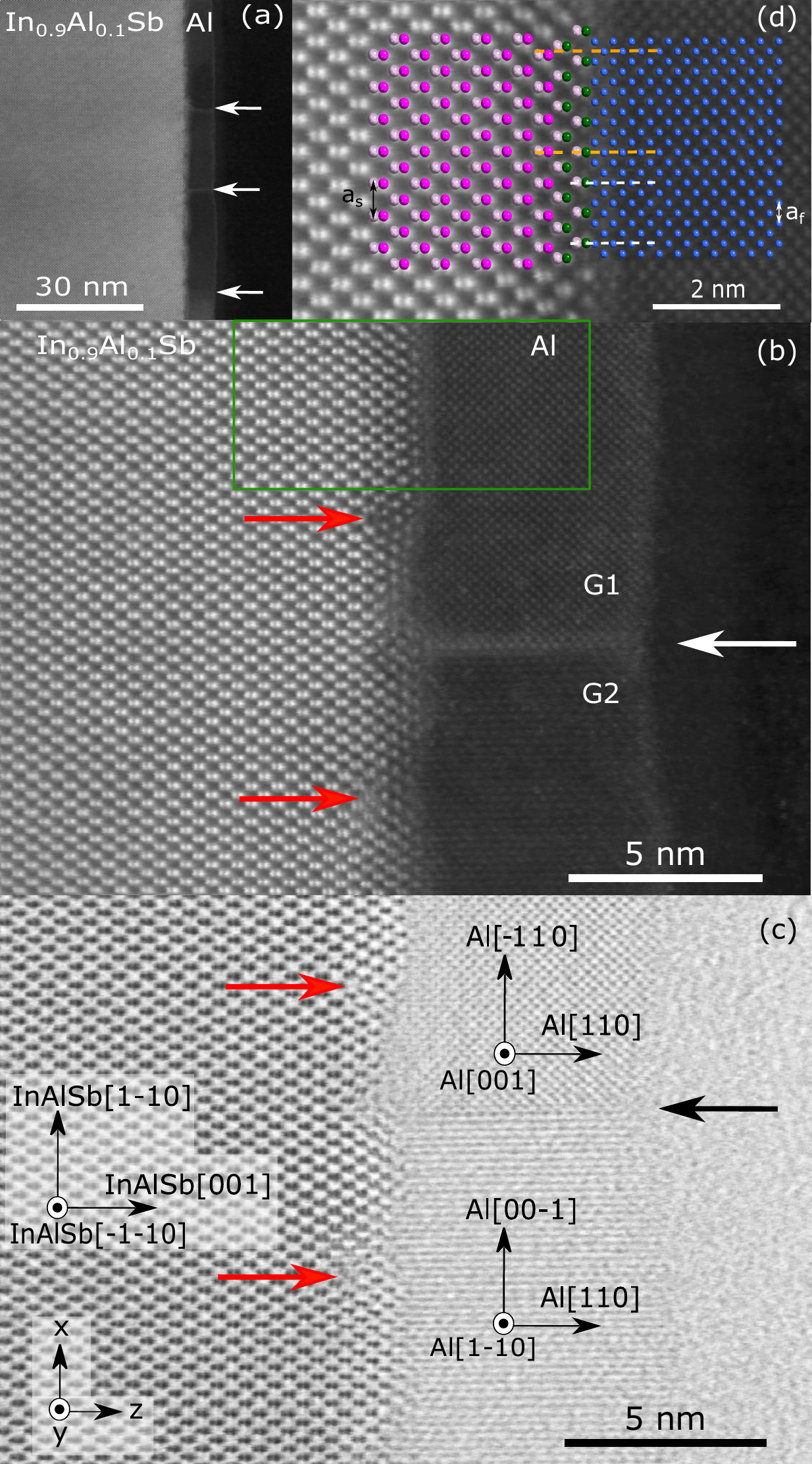} 
\caption{ STEM characterization of the Al-In$_{0.9}$Al$_{0.1}$Sb interface of sample C.  (a) HAADF STEM image, highlighting the formation of Al grains. White arrows indicate grain boundaries visible with light contrast. (b) High resolution HAADF and (c)  BF STEM images focusing on the interface. 
Sharp chemical interface is visible but presence of defects in semiconductor such as misfit dislocations is noticed (see red arrows and text for explanation).  Two Al grains (grain 1, G1, and grain 2, G2) are visible on (b) and (c), sharing the same growth orientation [110]. (d) Zoomed-in on the green frame of (b). Crystalline structure model, obtained with CrystalMaker software \citep{CrystalMaker}, overlaps the HRSTEM image. Good agreement is found  between the data and the model with strained In$_{0.9}$Al$_{0.1}$Sb and 2ML of relaxed InAs. $a_f$ and $a_s$ define the Al film and the semiconductor lattice parameters along $x$-axis, respectively.  Al atomic columns are represented in dark blue, As ones in green, In-based ones in light purple and Sb ones in magenta. The two dashed white and orange lines are guide to the eyes to show the difference of domain matching between Al and InAs and Al and strained In$_{0.9}$Al$_{0.1}$Sb, respectively (see text for discussion). Axis $x,y,z$ are defined in (c).   }
\label{Figure4}
\end{figure*}

Figure \ref{Figure4} presents STEM imaging of sample C in the vicinity of Al-semiconductor interface. A uniform Al film is observed in Figure \ref{Figure4}(a) with around 20 to 30 nm wide grains separated by sharp boundaries (highlighted by white arrows) perpendicular to the interface.
The bright contrast of the grain boundaries can be due to Ga incorporation during FIB preparation process (EDX data not shown here).

Atomic resolution high angle annular dark field (HAADF) and bright field (BF) STEM images are presented in Figures \ref{Figure4}(b) and (c), respectively, with focus on the interface.
In the HAADF image of Figure \ref{Figure4}(b), intensities scale with the average atomic number and total number of atoms in each column. As a result, light Al atoms appear dark while the heavier In$_{0.9}$Al$_{0.1}$Sb compound is brighter. The interface between these two materials appears sharp with a noticeable decay of atomic contrast over 2-3 ML toward the semiconductor, which is associated to the presence of the epitaxial InAs interlayer. 

A few defects associated with misfit dislocations are identified close to the interface (marked by red arrows) at a distance of about 2-3 ML below the Al layer. The remote position of these defects compared to the interface with the Al layer and the atomic contrast surrounding these defects suggest that they appear at the initiation of the highly mismatched growth of InAs on In$_{0.9}$Al$_{0.1}$Sb. 
Figure \ref{Figure4}(d) zooms on the Al-semiconductor interface, corresponding to the green frame of Figure \ref{Figure4}(b). The model of crystalline structure is produced by CrystalMaker software \citep{CrystalMaker} assuming a relaxed InAs interlayer on top of strained In$_{0.9}$Al$_{0.1}$Sb. The actual level of relaxation of the InAs interlayer can't be determined from this analysis.  Misfit dislocations at the InAs-In$_{0.9}$Al$_{0.1}$Sb interface potentially drive the relaxation. 
From the data presented here and additional micrographs (see Section C of Supplementary Information), the misfit dislocations present at InAs-In$_{0.9}$Al$_{0.1}$Sb interface and the associated relaxation do not seem to be correlated to the position of the grain boundaries in the Al film. 

The In$_{0.9}$Al$_{0.1}$Sb semiconductor exhibits clear dumbbells associated with III and V element atomic columns. They are regularly distributed showing no evidence of structural defects in this layer.  Clear atomic columns are also observable in the Al layer, specifically for the first grain, identified as G1 on Figure \ref{Figure4}(b). In the second grain (labeled G2 on Figure \ref{Figure4}(b)), atomic columns are not distinguishable but we can see lattice planes perpendicular to the interface, suggesting that this grain orientation differs from G1 and is slightly misaligned with the zone axis. To determine the crystalline orientation of each of these grains, we have studied the reciprocal space patterns obtained from local fast Fourier transforms (FFT) (see Section B of Supplementary Information). Epitaxial relationships between Al grains and the semiconductor have been deduced from these patterns and are indicated in Figure \ref{Figure4}(c) for each grain. The two grains share the same [110] growth orientation (along $z$ axis, defined in Figure \ref{Figure4}(c)). However, the in-plane orientations along $x$ and $y$ axis (axis defined in Figure \ref{Figure4}(c))  differ and indicate a rotation of about 90 $\degree$ in the $(x,y)$ plane between G1 and G2. 
Additional HRSTEM micrographs acquired at different positions of the lamella confirm the predominance of [110] growth orientation for Al grains (see Section C of  Supplementary Information). Evidence of grains with additional tilt in $(x,z)$ and $(y,z)$ planes was found but the density appears low.

As highlighted in Figure \ref{Figure4}(d) by the two white dashed lines, the growth of Al (110) on top of the relaxed InAs (001) screening layer allows minimization of the in-plane domain mismatch with the semiconductor below in comparison to the direct deposition of Al on InSb or strained In$_{0.9}$Al$_{0.1}$Sb (see orange dashed lines). Along $x$ axis, we can see that 3 lattice parameters of Al along [-110] match with 2 of InAs along [1-10], corresponding to a mismatch that can be as low as 
($\frac{3 Al[-110]}{2 InAs [1-10]}, 0.25 \%$) if InAs is fully relaxed. Concurrently, along $y$ axis, we estimate a higher domain mismatch with ($\frac{1 Al [001]}{1 InAs [-1-10]}, -5.5 \%$). For Al directly deposited on InSb or strained In$_{0.9}$Al$_{0.1}$Sb, the domain along $x$ is larger to minimize the mismatch and consists of 5 lattice parameters of Al in front of 3 of In$_{0.9}$Al$_{0.1}$Sb, as highlighted by the orange dashed lines in Figure \ref{Figure4}(d). Under these conditions, the domain mismatches are much larger than on InAs with  ($\frac{5 Al[-110]}{3 InAlSb [1-10]}, 4.1 \%$) and ($\frac{1 Al [001]}{1 InAlSb [-1-10]}, -11.6 \%$) along $x$ and $y$ axis, respectively. Understanding the relation between the Al-semiconductor domain mismatch and the formation of grains is beyond the scope of this paper.

\section{Conclusion}
In conclusion, we have shown that the insertion of a Sb-free screening layer of 2ML InAs in between In$_{x}$Al$_{1-x}$Sb, with $x \geq 0.9$, and Al enables to significantly enhance the durability of these hybrid structures. Optimization of sample preparation process has made possible the analysis of these structures with HRSTEM techniques. A reduction of the domain mismatch between Al and  the semiconductor underneath is observed with the relaxation of the InAs interlayer.

\section{Acknowledgments}
This work was supported by Microsoft Quantum.

\section{Supplementary information}

\subsection{Additional X-ray Reflectivity data}
\begin{figure*}[h]
\includegraphics[scale=0.4]{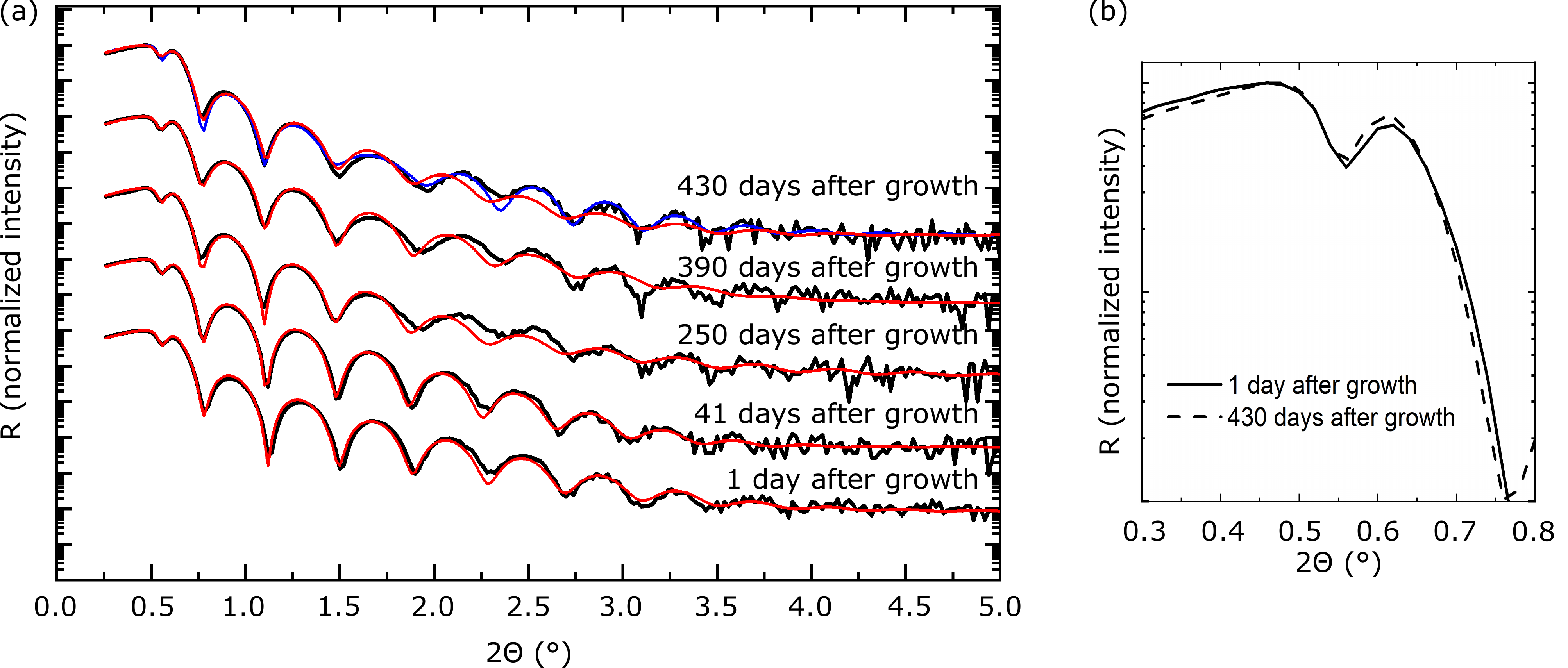} 
\caption{(a) XRR spectra with normalized reflectivity R as a function of $2\theta$  for sample B. The different spectra have been measured 1, 41, 250, 390 and 430 days after growth and are shifted for clarity.  
The red curves correspond to the fits using X'pert Reflectivity \citep{Reflectivity} with 19.8 $\pm$ 0.2 nm of Al and 1.7 $\pm$ 0.3 nm of Al$_2$O$_3$. The blue curve is a fit considering a 2 nm AlSb layer at the superconductor-semiconductor interface. (b) Zoom-in at low angle of the XRR spectra measured 1 and 430 days after growth, showing no shift of critical angle. }
\label{FigureS0}
\end{figure*}

Additional XRR spectra for sample B are presented in Figure \ref{FigureS0}(a) for 1, 41, 250, 390 and 430 days after growth. 
All these spectra are identical at low angles, as emphasized in Figure \ref{FigureS0}(b) with no shift of critical angle, demonstrating that the top layer is still mostly composed of pure Al more than 430 days after growth. 
These spectra have all been fitted with 19.8 $\pm$ 0.2 nm of Al and 1.7 $\pm$ 0.3 nm of Al$_2$O$_3$ (see red curves). One can notice a slight shift between the experimental data and the fits for spectra acquired more than 250 days after growth around $2 \Theta = 2.0 - 2.5 \degree$. The incorporation of AlSb or AlInSb at the superconductor-semiconductor interface in the fit model does not improve the fit for the spectra measured 250 and 390 days after growth.
On the contrary, for the spectrum acquired 430 days after growth, adding 2 nm of AlSb at the interface in the fit model allows to slightly better match the oscillations (see blue curve). From these data, we can conclude that there is no  evidence of a reaction between the Al film and the InSb layer more than 390 days after growth. The Al layer appears pure. After 430 days, the fit in blue might suggest a beginning of reaction at the superconductor-semiconductor interface with 2 nm of AlSb, but most of the Al layer is still pure. 
It is worth noting that improving the storage conditions of these samples can further increase their lifetime with for instance the use of a refrigerated dry box.

\newpage

\subsection{Local Fast Fourier Transform of grains 1 and 2}

\begin{figure}[h]
\includegraphics[scale=0.7]{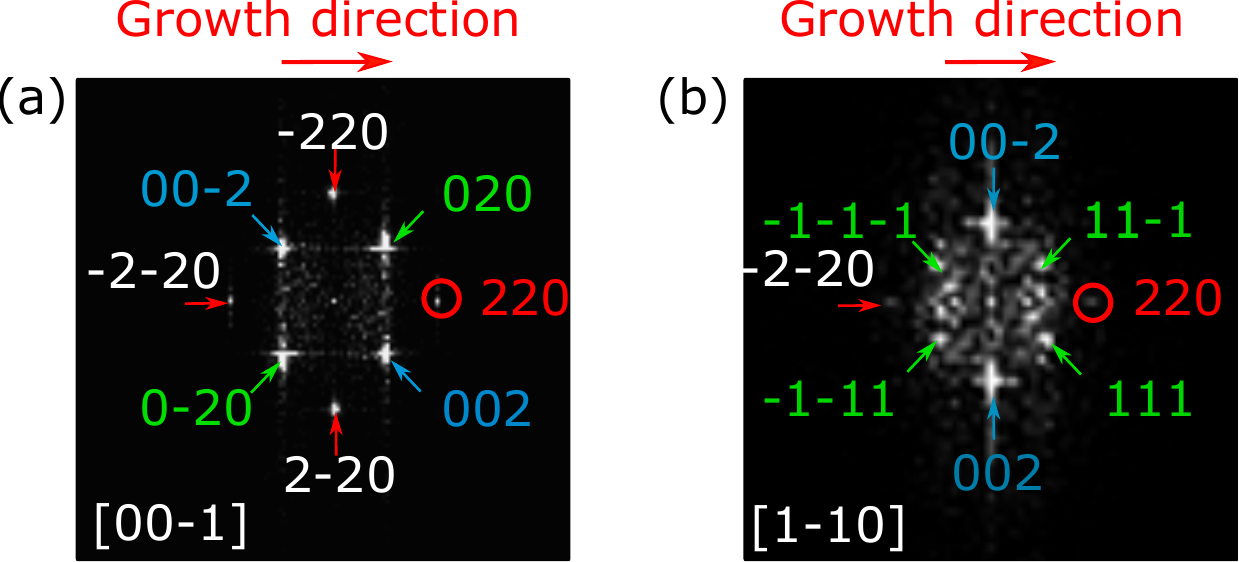} 
\caption{Local Fast Fourier Transform (FFT) performed on grain 1 (a) and grain 2 (b) of Figure \ref{Figure4}. 220 crystalline orientation is parallel to the growth direction for both grains, indicating [110] growth direction for the two grains. The zone axis is indicated in the bottom left for each diffraction pattern. }
\label{FigureS1}
\end{figure}

Figures \ref{FigureS1}(a) and (b) present local FFTs performed on Al grain 1 and grain 2 of Figure \ref{Figure4}, respectively. These FFTs allow to access the reciprocal space patterns  for each grain. 
The labeled crystalline orientations are identified by measuring the distance between the different spots and comparing them to simulated diffraction patterns obtained via SingleCrystal software \citep{CrystalMaker}. 

The FFT of grain 1 (see Figure \ref{FigureS1}(a)) is characterized by clear spots and high signal to noise ratio, thanks to the good alignment of grain 1 with the zone axis allowing to observe clear atomic columns in Figure \ref{Figure4}. The measured distances between outer spots (see red arrows) indicate $\left\{220\right\}$  planes, while the inner spots (see green and blue arrows) are identified to belong to both $\left\{002\right\}$ and $\left\{020\right\}$  planes. The growth direction of grain 1 is thus along [110]. The epitaxial relationship of grain 1 with semiconductor is indicated in Figure \ref{Figure4}(c). 

The FFT of grain 2 (see Figure \ref{FigureS1}(b)) has a lower signal to noise ratio that is explained by a slight misalignment of the grain with the zone axis. The distance between outer horizontally aligned spots (see red arrows) is consistent with $\left\{220\right\}$  planes, while the distance between the outer vertically aligned spots (see blue arrows) is, contrary to grain 1, in agreement with $\left\{002\right\}$  planes. The inner spots (see green arrows) are identified to belong to $\left\{111\right\}$ planes. This difference of pattern compared to grain 1 suggests a rotation in the plane $(x,y)$ of about $90 \degree$ between the two grains. Epitaxial relationship of grain 2 with semiconductor is indicated in Figure \ref{Figure4}(c).

\newpage

\subsection{Additional HRSTEM imaging}
\begin{figure*}[h]
\includegraphics[scale=0.9]{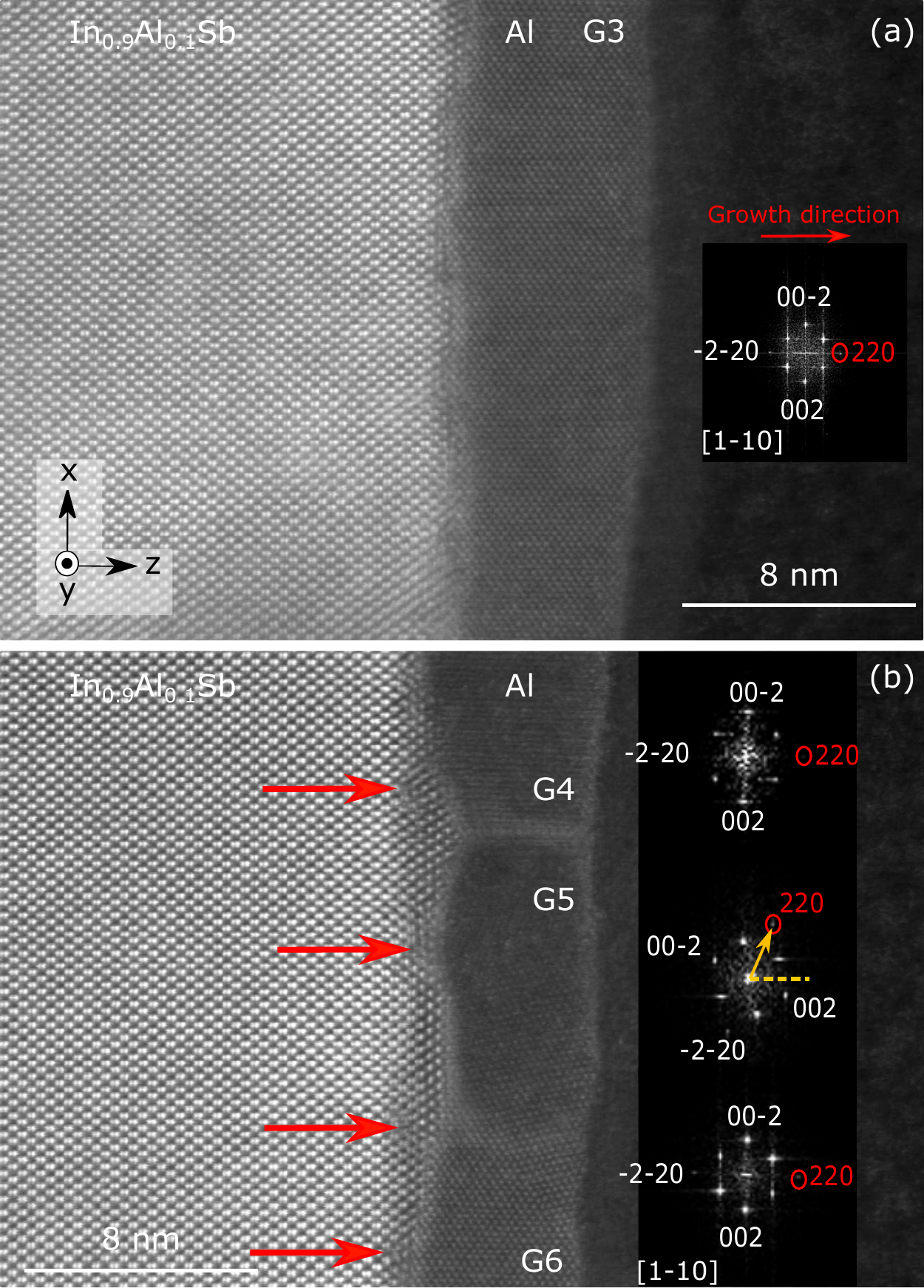} 
\caption{ HRSTEM characterization at different locations of the lamella in HAADF mode. (a) shows the presence of a unique grain (G3) with [110] growth orientation, while (b) shows the presence of three grains (G4, G5 and G6). G4 and G6 have [110] growth orientation while G5 is tilted  by $\sim 65 \degree$ in the $(x,z)$ plane (see tilt of atomic columns and the yellow arrow on FFT pattern).  FFT patterns are presented on the right for each grain. Red arrows indicate the presence of misfit dislocations. Axis $x,y,z$ are defined in (a).  }
\label{FigureS2}
\end{figure*}

Additional HAADF HRSTEM images of sample C are presented in Figure \ref{FigureS2}. 
Fig. \ref{FigureS2}(a) displays a unique Al grain, denoted as G3, more than 20 nm wide, while three grains G4, G5 and G6 are visible on Fig. \ref{FigureS2}(b) on the same length scale.  The  presence of misfit dislocations at the InAs-In$_{0.9}$Al$_{0.1}$Sb interface is  observed (see red arrows in Fig. \ref{FigureS2}(b)). They do not match the position of the grain boundaries, thus supporting the absence of correlation between these two features. 

Local FFTs have been performed on each of the presented Al grains. A similar pattern, than observed for grain 2 of the main text, is evidenced for all of these grains. Al growth direction is along [110] with in-plane direction $x$ along [00-1]. 
When looking closely at the middle grain, labeled as G5 in Figure \ref{FigureS2}(b), one can see atomic columns which are clearly tilted compared to those of the adjacent grain G6. This tilt appears also on the FFT pattern and we can estimate a rotation of $\sim 65 \degree$ in the $(x,z)$ plane between the 220 reflections of the two grains (see yellow arrow). 
G4 and G6 have the same FFT pattern but one can notice that the FFT signal to noise ratio is different. Concurrently, G6 shows clear atomic columns while only lattice planes are visible for G4. These differences can be explained by a slight tilt between these two grains.

\end{document}